\documentclass[aps,prb,manuscript]{revtex4-1}
\usepackage{graphicx}
\usepackage{epsfig}
\usepackage{color}
\usepackage{bm}

\begin{document}
\draft
\title{Quantum-Electron Back Action on Hybridization of Radiative and Evanescent Field Modes}

\author{Andrii Iurov$^{1}$, Danhong Huang$^{2}$, Godfrey Gumbs$^{3}$, Wei Pan$^{4}$ and A. A. Maradudin$^{5}$}
\affiliation{$^{1}$Center for High Technology Materials, University of New Mexico, 1313 Goddard SE, Albuquerque, New Mexico, 87106, USA\\
$^{2}$Air Force Research Laboratory, Space Vehicles Directorate, Kirtland Air Force Base, New Mexico 87117, USA\\
$^{3}$Department of Physics and Astronomy, Hunter College of the City University of New York, 695 Park Avenue New York, New York 10065, USA\\
$^{4}$Sandia National Laboratories, Albuquerque, New Mexico 87185, USA\\
$^{5}$Department of Physics and Astronomy, University of California, Irvine, California 92697, USA}

\date{\today}

\begin{abstract}
A back action from Dirac electrons in graphene on the hybridization of radiative and evanescent fields is found as an analogy to Newton's third law.
Here, the back action appears as a localized polarization field which greatly modifies an incident surface-plasmon-polariton (SPP) field.  
This yields a high sensitivity to local dielectric environments and provides a scrutiny tool for molecules or proteins selectively bounded with carbons.  
A scattering matrix is shown with varied frequencies nearby the surface-plasmon (SP) resonance for the increase, decrease and even a full suppression of the polarization field, 
which enables accurate effective-medium theories to be constructed for Maxwell-equation finite-difference time-domain methods.
Moreover, double peaks in the absorption spectra for hybrid SP and graphene-plasmon modes are significant only with a large conductor plasma frequency, but are overshadowed by
a round SPP peak at a small plasma frequency as the graphene is placed close to conductor surface. 
These resonant absorptions facilitate the polariton-only excitations, leading to polariton condensation for a threshold-free laser. 
\end{abstract}
\pacs{PACS:}
\maketitle

\noindent
\underline{\em Introduction}:\ \ ``For every action, there is an equal and opposite reaction.'' as famously stated by the Newton's third law in classical mechanics. It is known that 
when light is incident on a semiconductor, its energetic photons can lift electrons from a lower 
valence band to a higher conduction band, leaving many electron-hole pairs in the system\,\cite{add8,add26}. Simultaneously, its electric-field component will further push away these negatively (positively) charged electrons (holes) 
in opposite field directions. 
For this instance, however, \textsl{do excited electrons and holes exert an action back on the incident light?} The answer to this question lies in 
the induced optical-polarization field as a collection of local dipole moments from 
many displaced electrons and holes\,\cite{add9,add29}, which plays a role in scattering the electric-field component of the incident light\,\cite{ref1,add12}.  
Therefore, the quantum nature of Dirac electrons\,\cite{add1,add2,add3,add4} is expected to be reflected in this electron back action on the incident light with a complex distribution of Landau-damping regions 
in comparison with that of two-dimensional electron gases in a quantum well.
\medskip

\begin{figure}[p]
\centering
\includegraphics[width=0.55\textwidth]{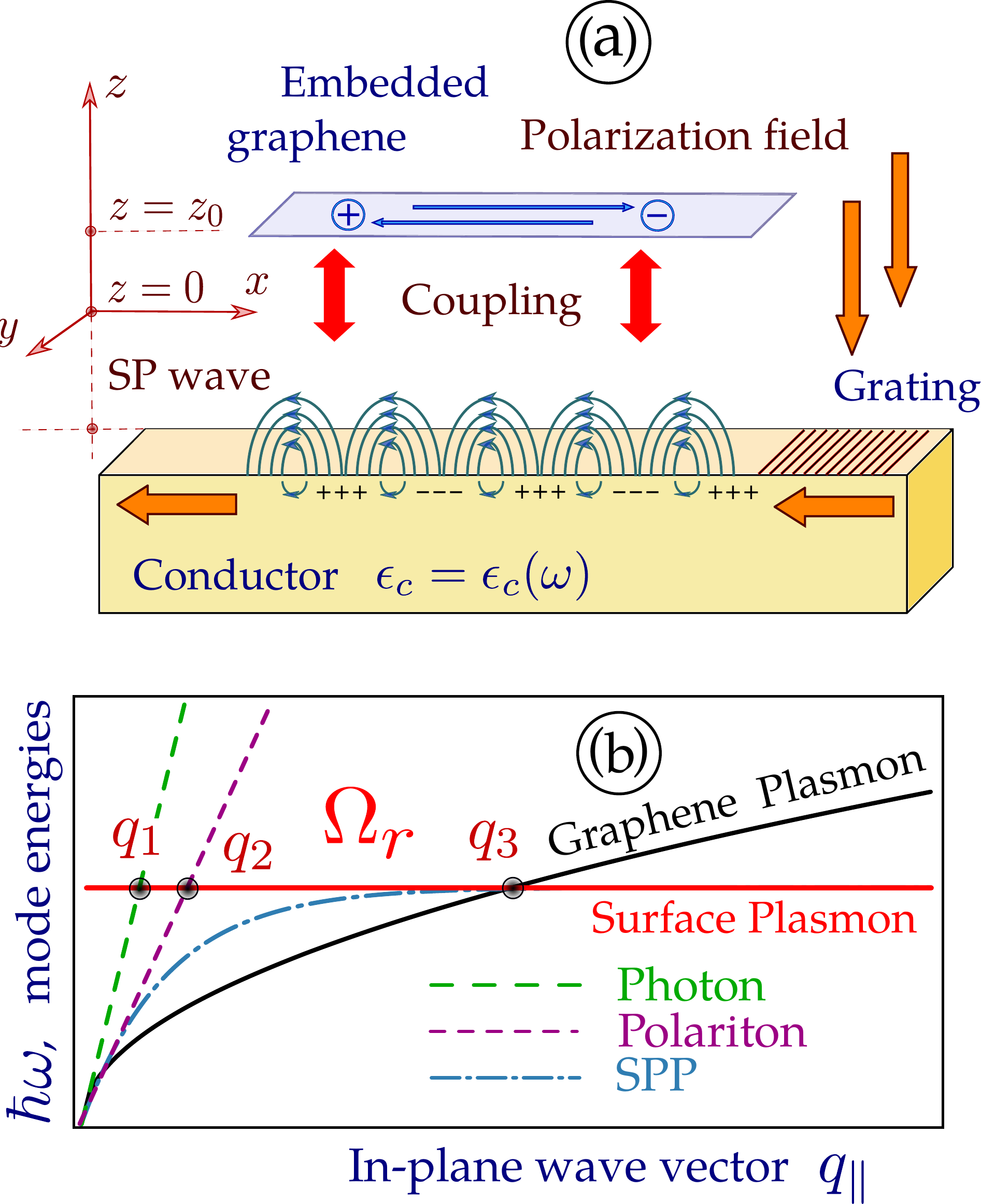}
\caption{Schematic ($a$) for a thick conductor $\epsilon_{\rm c}(\omega)$ and an embedded graphene sheet at $z=z_0$ above its surface at $z=0$ within a cladding dielectric $\epsilon_{\rm d}$.
The surface-plasmon-polariton field (SPPF) is excited by light incident on a grating. The
propagating SPPF excites Dirac electrons in graphene, and the induced graphene polarization modifies the SPPF by resonant scattering. 
Illustration ($b$) for energy dispersions of photons, polaritons, surface-plasmon polaritons (SPPs), graphene plasmons (G-Ps) and surface plasmons (SPs), 
where three labeled circles indicate the mode hybridizations.}
\label{f1}
\end{figure}

For a hybrid system illustrated in Fig.\,\ref{f1}$a$, we encounter radiative field modes, such as photons and polaritons\,\cite{oe1,oe2,oe3,oe4,add28}, 
as well as evanescent field modes, e.g., surface and graphene plasmons\,\cite{add5,add6,add7}. 
Researches on optical responses of graphene electrons have been reported before\,\cite{add7,add10}, but most of those efforts are limited to the radiation or grating-deflection field coupling. 
In contrast to the plane-wave-like external field, we explore the surface-plasmon-polariton near-field\,\cite{add11,add15,add27} coupling to graphene electrons with a different dispersion relation from the usual
linear one, i.e., $\omega=qc$, for the free-space light. In our case, the graphene sheet is brought very close to the surface of a conducting substrate so that the hybridization of radiative and evanescent fields can 
occur\,\cite{add24}. As a result, the non-dispersive surface-plasmon mode is able to hybridize successively with radiative photon and polariton modes\,\cite{oe1,oe2}, 
as well as with the spatially-localized graphene plasmon mode\,\cite{add14}, as shown schematically in Fig.\,\ref{f1}$b$. 
\medskip

Such a distinctive dispersion relation of the hybrid quantum-plasmon modes should be experimentally observable in optical spectra\,\cite{add18,add25,add31,add32}.
The effective scattering matrix\,\cite{add19,add30} from such a coupled system is found significantly different from either the graphene sheet or the conductor and it displays
distinctive features from the retarded longitudinal Coulomb interaction\,\cite{add12} between electrons in the graphene sheet and conductor.
This scattering matrix can be employed for constructing an effective-medium
theory\,\cite{add20,add21,add22,add23} to study the optical properties of inserted biomolecules and metamaterials between the graphene sheets and the conductor surface. Therefore, a locally environment-sensitive super-resolution
near-field imaging can be developed for functionalized biomolecules bounded with metallic nanodots and nanorods or even carbon atoms of graphene\,\cite{add13}.
\medskip

\noindent 
\underline{\em Theory and Methods}:\ \ As illustrated in Fig.\,\ref{f1}$a$, our model system consists a thick conductor and a dielectric-embedded graphene sheet above its surface.
A surface-plasmon-polariton field (SPPF) can be excited by incident light on a surface grating. This surface-propagating SPPF couples to
Dirac electrons in graphene, and the induced polarization field from graphene acts back simultaneously on the SPPF as a resonant scatter.
\medskip

Using the Green's function approach,\,\cite{ref1} we convert Maxwell's equation for the electric field $\mbox{\boldmath$E$}({\bf r},\omega)$
into an integral equation in the spatial (${\bf r}$) domain, including a nonlocal source term to scatter the incident SPPF $\mbox{\boldmath${\cal E}$}^{\rm inc}({\bf r},\omega)$, where $\omega$ is the light frequency.
After a Fourier transform of this integral equation with respect to ${\bf r}_\|$, we obtain ($\mu,\nu=1,2,3$)

\begin{equation}
E_{\mu}({\bf q}_\|,\omega\vert x_3)={\cal E}^{\rm inc}_{\mu}({\bf q}_\|,\omega\vert x_3)
+\frac{\omega^2}{\epsilon_0c^2}\sum\limits_{\nu}
g_{\mu\nu}({\bf q}_\|,\omega\vert x_3,z_0)\,{\cal P}_{\nu}^{\rm s}({\bf q}_\|,\omega)\ ,
\label{eq1}
\end{equation}
where ${\bf r}=\{{\bf r}_\|,x_3\}$, $x_3=z_0$ denotes the graphene-sheet position, and
$g_{\mu\nu}({\bf q}_\|,\omega\vert x_3,z_0)$ is the Fourier transformed Green's function matrix\,\cite{add12} which corresponds to a retarded coupling between graphene electrons to the incident SPPF. 
Based on linear-response theory\,\cite{ref2}, we find the graphene polarization field in Eq.\,(\ref{eq1})
${\cal P}_{\nu}^{\rm s}({\bf q}_\|,\omega)=\epsilon_0\,\chi^{(0)}_{\rm s}(q_\|,\omega)\left(1-\delta_{\nu 3}\right)E_{\nu}({\bf q}_\|,\omega\vert z_0)$,
where $\chi^{(0)}_{\rm s}(q_\|,\omega)=e^2\,\Pi^{(0)}_{\rm s}(q_\|,\omega)/[\epsilon_0(q_\|^2-\epsilon_{\rm d}\,\omega^2/c^2)]$ represents the optical-response function\,\cite{ref3}, and 
$\Pi^{(0)}_{\rm s}(q_\|,\omega)$ is the density-density correlation function of graphene electrons\,\cite{add16}.
Setting $x_3=z_0$ in Eq.\,(\ref{eq1}), we obtain a self-consistent equation for $\mbox{\boldmath$E$}({\bf q}_\|,\omega\vert z_0)$.
Furthermore, if $\mbox{\boldmath${\cal E}$}^{\rm inc}({\bf q}_{\|},\omega\vert z_0)=0$ is assumed in this self-consistent equation, we get the dispersion equation for the hybrid plasmon modes, 
and the resulting dispersion relation $\omega=\Omega({\bf q}_\|\vert z_0)$, as illustrated in Fig.\,\ref{f1}$b$, is determined from the real part of the secular equation 
${\cal D}et\{\tensor{\mbox{\boldmath${\cal C}$}}({\bf q}_\|,\omega\vert z_0)\}=0$, where $\tensor{\mbox{\boldmath${\cal C}$}}({\bf q}_\|,\omega\vert z_0)=\delta_{\mu\nu}-(\omega/c)^2\,
g_{\mu\nu}({\bf q}_\|,\omega\vert z_0,z_0)\left(1-\delta_{\nu 3}\right)\chi^{(0)}_{\rm s}(q_\|,\omega)$ is the complex coefficient matrix.
After calculating the inverse of $\tensor{\mbox{\boldmath${\cal C}$}}$, $\mbox{\boldmath$E$}({\bf r}_\|,\omega\vert x_3)$ can be expressed explicitly as

\[
E_{\mu}({\bf r}_\|,\omega\vert x_3)={\cal E}^{\rm inc}_{\mu}({\bf r}_\|,\omega\vert x_3)+\frac{\omega^2}{c^2}\int \frac{d^2{\bf q}_{\|}}{(2\pi)^2}\,
e^{i{\bf q}_\|\cdot{\bf r}_\|}\,\chi^{(0)}_{\rm s}(q_\|,\,\omega)
\]
\begin{equation}
\times\sum\limits_{\nu}\left\{g_{\mu\nu}({\bf q}_\|,\omega\vert x_3,z_0)\left(1-\delta_{\nu 3}\right)
\left[\sum_{\mu^\prime}\,{\cal C}^{-1}_{\nu\mu^\prime}({\bf q}_\|,\omega\vert z_0)\,{\cal E}^{\rm inc}_{\mu^\prime}({\bf q}_\|,\omega\vert z_0)\right]\right\}\ .
\label{eq2}
\end{equation}
From Eq.\,(\ref{eq2}) the Fourier transformed scattering matrix\,\cite{add19} is easily found to be

\begin{equation}
\alpha_{\mu\nu}^{\rm eff}({\bf q}_\|,\omega\vert x_3)=\frac{\omega^2}{c^2}\,\chi^{(0)}_{\rm s}(q_\|,\omega)\,
\sum\limits_{\nu^\prime}\,g_{\mu\nu^\prime}({\bf q}_\|,\omega\vert x_3,z_0)\left(1-\delta_{\nu^\prime 3}\right)
{\cal C}^{-1}_{\nu^\prime\nu}({\bf q}_\|,\omega\vert z_0)\ .
\label{eq3}
\end{equation}
Additionally, by using the calculated $\mbox{\boldmath$E$}({\bf r}_\|,\omega\vert z_0)$ in Eq.\,(\ref{eq2}), the absorption coefficient
$\beta_{\rm abs}(\omega\vert z_0)$ for the SPPF can be calculated\,\cite{add11,lorentz} based on the Lorentz function $\alpha_{\rm L}(\omega\vert z_0)$ given by

\[
\alpha_{\rm L}(\omega\vert z_0)=\frac{c}{\omega}
\left|\sum_{\mu,\nu}\,\hat{e}_{\mu}\,{\cal C}^{-1}_{\mu\nu}(\mbox{\boldmath$k$}_0,\omega\vert z_0)\left[i\hat{k}_\nu\beta_3(\omega)-\hat{x}_\nu k_0(\omega)\right]e^{-\beta_3(\omega)z_0}\right|
\]
\begin{equation}
\times\left(\frac{2\pi e^2}{\epsilon_0\epsilon_{\rm r}k_0}\right)\sqrt{1-\frac{\epsilon_{\rm d}\omega^2}{k_0^2c^2}}
\left\{\Pi^{(0)}_{\rm s}(k_0,\omega)+[\Pi^{(0)}_{\rm s}(k_0,-\omega)]^\ast\right\}\ ,
\label{eq4}
\end{equation}
where $\epsilon_{\rm r}$ is the graphene dielectric constant,
$\hat{\mbox{\boldmath$e$}}$ and $\hat{\mbox{\boldmath$k$}}$ are the unit polarization and wave vectors of the SPPF, $\hat{\mbox{\boldmath$x$}}=(0,0,1)$ is the spatial unit vector, 
$\mbox{\boldmath$k$}_0\equiv{\rm Re}[k_0(\omega)]\hat{\mbox{\boldmath$k$}}$, $\beta_3(\omega)=\sqrt{k_0^2(\omega)-\omega^2/c^2}$, 
$k_0(\omega)=(\omega/c)\sqrt{\epsilon_{\rm d}\epsilon_{\rm c}(\omega)/[\epsilon_{\rm d}+\epsilon_{\rm c}(\omega)]}$, $\epsilon_{\rm d}$ is the cladding-layer dielectric constant,
and $\epsilon_{\rm c}(\omega)=\epsilon_{\rm s}-\Omega_p^2/\omega^2$ for the conductor.
\medskip

\noindent
\underline{\em Results and Discussions}:\ \ In our numerical calculations, 
we use the Fermi wave vector $k_{\rm F}=\sqrt{\pi n_0}$ as the scale for wave numbers, $1/k_{\rm F}$ for lengths, and $E_{\rm F}=\hbar v_{\rm F}k_{\rm F}$ for energies. 
The direction of SPPF propagation is chosen $\hat{\mbox{\boldmath$k$}}=(1,0,0)$ for simplicity, and we also set $\epsilon_{\rm s}=13.3$, $\epsilon_{\rm d}=\epsilon_{\rm r}=2.4$, 
$v_{\rm F}=1\times 10^8\,$cm/s, and $n_0=5\times 10^{11}\,$cm$^{-2}$ for the doping density in graphene.
Moreover, the half gap $\Delta=0$ is assumed unless it is stated in figure captions, and the resonant frequency $\Omega_r=\Omega_p/\sqrt{\epsilon_{\rm s}+\epsilon_{\rm d}}$ will be given directly in figure captions.
\medskip

\begin{figure}
\centering
\includegraphics[width=0.65\textwidth]{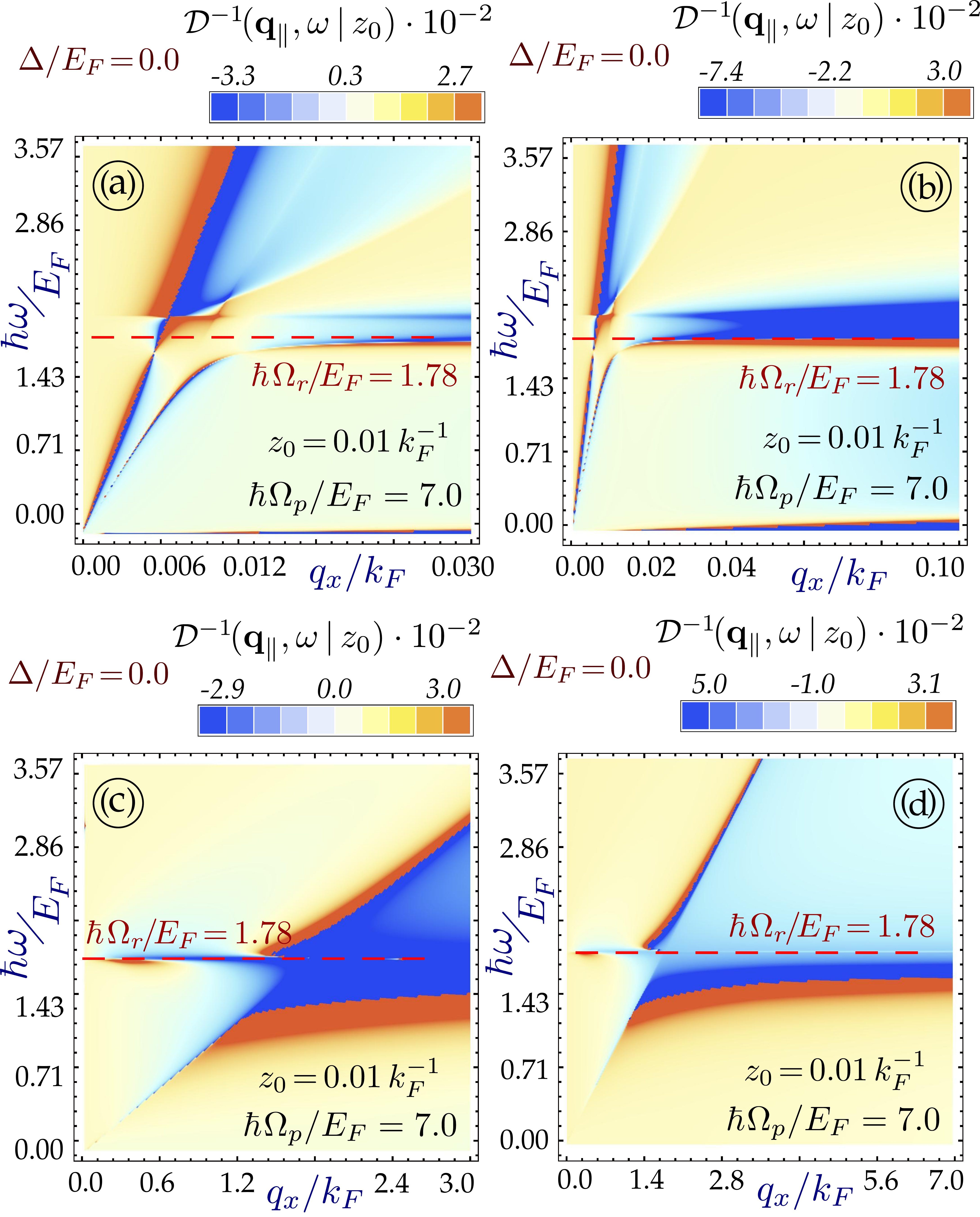}
\caption{Density plots for the real part of ${\cal D}^{-1}(q_x,\omega\vert z_0)\equiv 1/{\cal D}et\{\tensor{\mbox{\boldmath${\cal C}$}}(q_x,\omega\vert z_0)\}$ in four different $q_x$ ranges
growing from $0.03$ to $7$ with hybrid-plasmon dispersions indicated by jumps between positive (red) and negative (blue) peaks. Here, $k_Fz_0 = 0.01$, $\hbar\Omega_r/E_F=1.78$.}
\label{f2}
\end{figure}

For a retarded interaction between light and graphene electrons, both radiative and evanescent modes must be considered in the hybrid system. The radiative modes include photons and polaritons, while the 
evanescent modes appear as surface-plasmon polaritons (SPPs), graphene plasmons (G-Ps), and surface plasmons (SPs).
Figure\ \ref{f2} displays the real part of ${\cal D}^{-1}(q_x,\omega\vert z_0)\equiv 1/{\cal D}et[\tensor{\mbox{\boldmath${\cal C}$}}(q_x,\omega\vert z_0)]$ for four different $q_x$ ranges.
As seen from Fig.\,\ref{f2}$a$, in addition to the SPP mode,
the hybridizations of both radiative photon and polariton modes with localized SPs (labeled by $q_1$ and $q_2$ in Fig.\,\ref{f1}$b$) show up in this very small $q_x$ range. 
As the $q_x$ range slightly expands in Fig.\,\ref{f2}$b$, the SPP mode in Fig.\,\ref{f2}$a$ is fully developed, which is accompanied by 
a G-P mode at very low energies. As the $q_x$ range further increases in Figs.\,\ref{f2}$c$ and \ref{f2}$d$, the G-P energy exceeds that of the
SP. Consequently, the anticrossing of G-Ps with SPPs (labeled by $q_3$ in Fig.\,\ref{f1}$b$) shows up. 
\medskip

\begin{figure}
\centering
\includegraphics[width=0.65\textwidth]{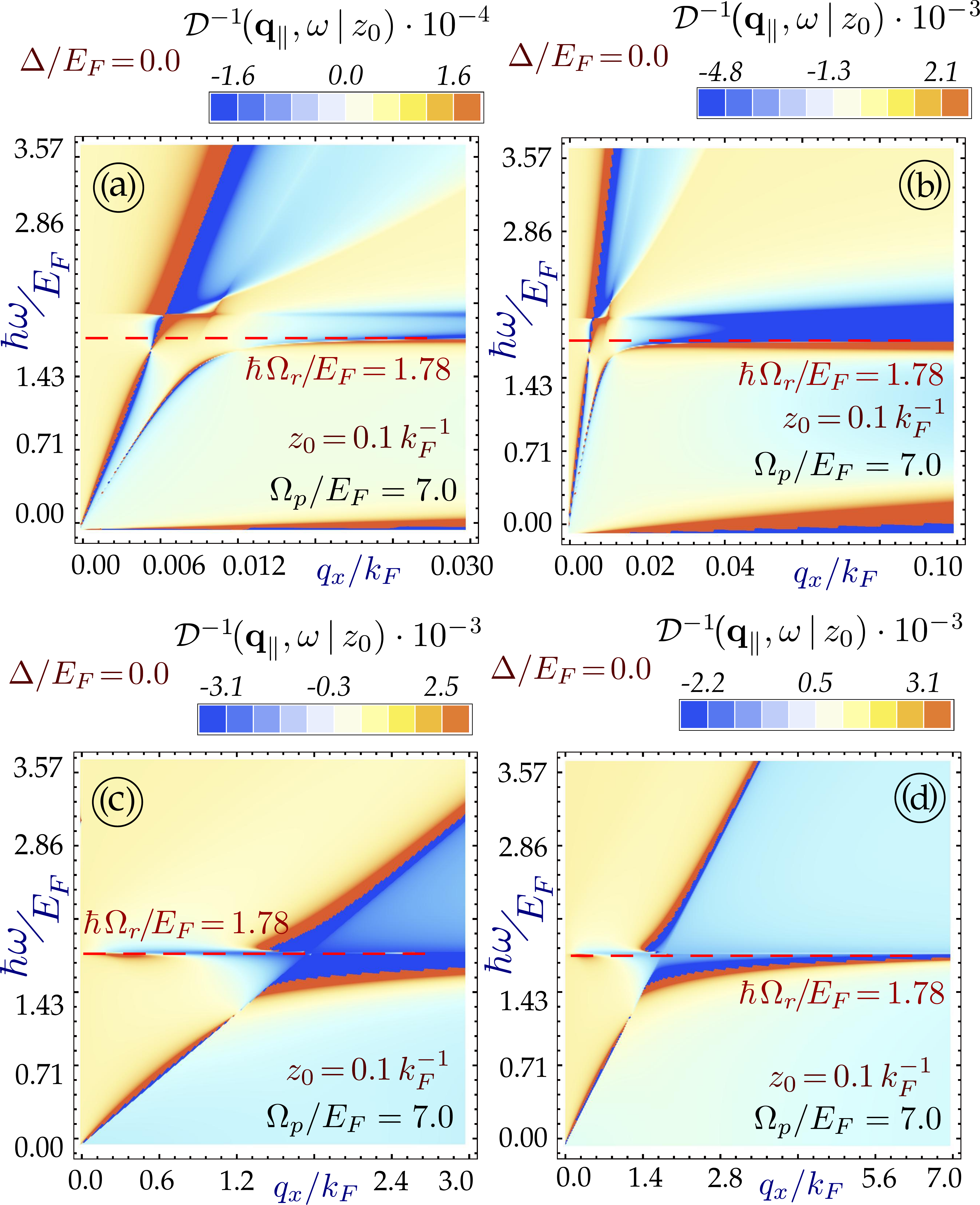}
\caption{Density plots for the real part of $\mathcal{D}^{-1}(q_x, \omega \, \vert z_0)$ for four different $q_x$ ranges growing from $0.03$ to $7$.
Here, $k_Fz_0 = 0.1$, $\hbar\Omega_r/E_F=1.78$.}
\label{f3}
\end{figure}

The $z_0$ dependence in the secular equation reflects the distinctive evanescent coupling between SPPs and G-Ps. Here,
the factor $g_{\mu\nu}(q_x,\omega\vert z_0,z_0)$ plays the role of a retarded SPP coupling to a spatially-separated G-P, while $\chi^{(0)}_{\rm s}(q_x,\,\omega)$ corresponds to the G-P optical response.
Therefore, their product represents the hybrid Dirac-SPP modes. By moving the graphene sheet a bit further from the conductor surface in Fig.\,\ref{f3}, the anticrossing gap shrinks due to 
decreased retarded coupling. Simultaneously, the strengths of all the plasmon, polariton and photon modes increase by more than one order of magnitude due to loss suppression of these modes to the conductor. 
\medskip

\begin{figure}
\centering
\includegraphics[width=0.65\textwidth]{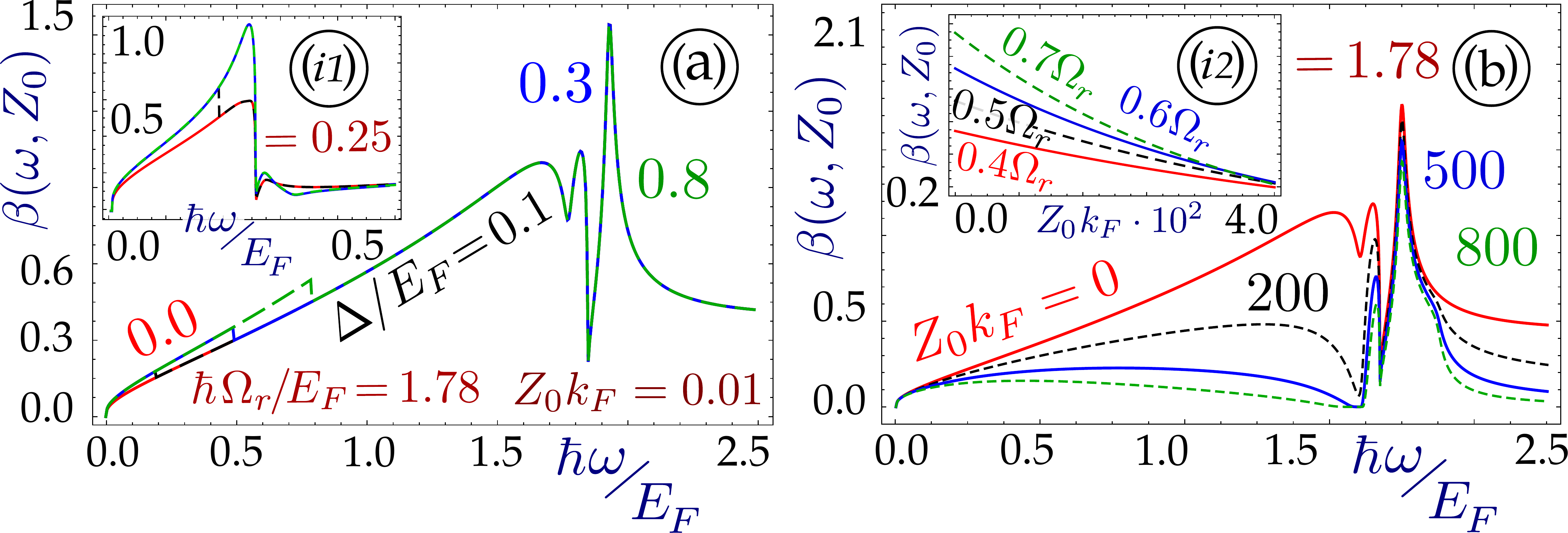}
\caption{Absorption spectra $\beta_{\rm abs}(\omega\vert z_0)$ displayed in ($a$) with $k_Fz_0=0.01$ for $\Delta/E_F=0,\,0.1,\,0.3,\,0.8$, 
where $\hbar\Omega_r/E_F= 1.78$ and $0.25$ in its inset. In ($b$),
$\beta_{\rm abs}(\omega\vert z_0)$ with $\hbar\Omega_r/E_F= 1.78$, $\Delta/E_F=0$ for 
$k_Fz_0=0,\,200,\,500,\,800$, and its $z_0$ dependence in the inset for $\omega/\Omega_r=0.4,\,0.5,\,0.6,\,0.7$.}
\label{f4}
\end{figure}

The incident SPPF suffers not only Ohmic loss during its propagation along the conductor surface, but also absorption loss by its coupling to G-Ps.  
Figure\ \ref{f4} presents the absorption spectra $\beta_{\rm abs}(\omega\vert z_0)$ with various $\Delta$ in ($a$) and different $z_0$ in ($b$).
From Fig.\,\ref{f4}$a$ we find three absorption peaks for $\Delta=0$, where two sharp ones correspond to G-Ps (right) and SPs (left), as seen from Figs.\,\ref{f2}$c$ and \ref{f2}$d$, 
with a deep dip between them for the opened anticrossing gap labeled by $q_3$ in Fig.\,\ref{f1}$b$. The lowest round peak is for SPP modes, as shown in 
Figs.\,\ref{f2}$a$ and \ref{f2}$b$, which is separated from the SP peak by a shallow dip labeled by $q_2$ in Fig.\,\ref{f1}$b$.
An absorption spike also shows up on the shoulder of this round peak for $\Delta>0$ and it shifts to higher energy with increasing $\Delta$. 
The spike feature for $\Delta>0$ is rooted in the light-frequency dependence in $k_0(\omega)$ for unique evanescent coupling and is a result of $\Delta$ driven modification 
in suppressing the Landau damping\,\cite{add16}, i.e. the ${\rm Im}[\Pi_{\rm s}^{(0)}(k_0,\omega)]$ term in Eq.\,(\ref{eq4}), for very large $k_0$ values 
with increasing $\omega$ towards $\Omega_r$. The absorption peak from the indistinguishable SPP and SP modes increases greatly in the inset of Fig.\,\ref{f4}$a$ for a lower SP resonance $\hbar\Omega_{\rm r}/E_{\rm F}=0.25$ since 
the SPPF decay is largely eased at a much smaller $q_x$. The enhancement of the SPP absorption peak is also seen in Fig.\,\ref{f4}$b$ for small $z_0$ due to reduced SPPF decay.
Additionally, we find from the inset of Fig.\,\ref{f4}$b$ that the decrease of $\beta_{\rm abs}(\omega\vert z_0)$ with increasing $z_0$ becomes much more dramatic as $\omega$ approaches $\Omega_r$ 
with increased SPPF localization. 
\medskip

\begin{figure}
\centering
\includegraphics[width=0.65\textwidth]{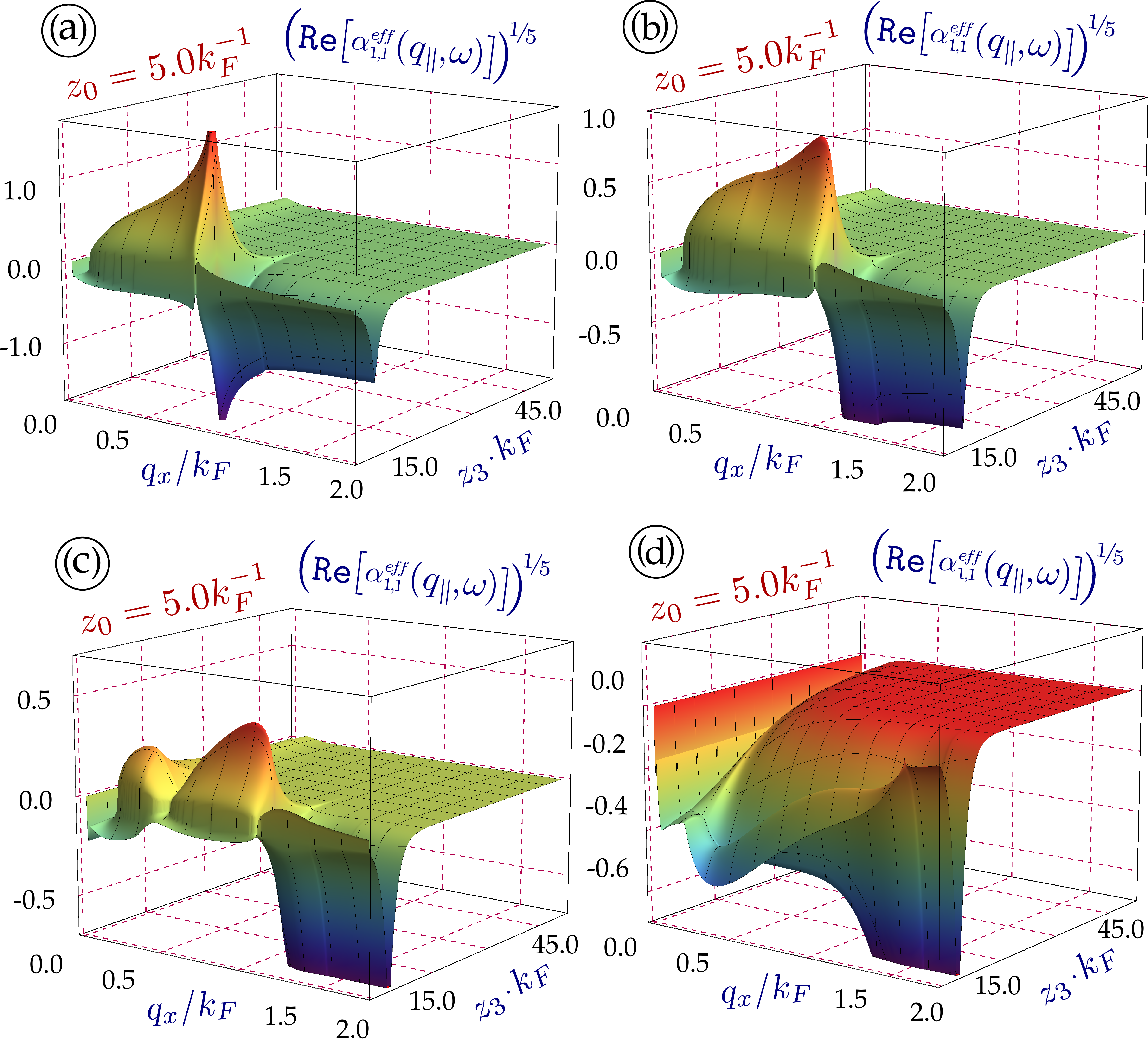}
\caption{3D plots for $[{\rm Re}\{\alpha^{\rm eff}_{11}(q_x,\omega\vert x_3)\}]^{1/5}$ with $\omega/\Omega_r=0.7\,(a),\,0.8\,(b),\,0.9\,(c),\,1.0\,(d)$, where $k_Fz_0 = 5$, $\hbar\Omega_r/E_F=1.78$.}
\label{f5}
\end{figure}

In addition to the SPPF optical absorption by G-Ps, a resonant scattering of the SPPF from G-Ps also happens, as described by Eq.\,(\ref{eq3}). Figure\ \ref{f5} presents 3D plots for 
$[{\rm Re}\{\alpha^{\rm eff}_{11}(q_x,\omega\vert x_3)\}]^{1/5}$ with four $\omega$ values, where the graphene sheet is put relatively close to the surface. Here, the scattering matrix is
$\alpha^{\rm eff}_{\mu\nu}\equiv\delta(E_{\mu}-{\cal E}_{\mu}^{\rm inc})/\delta{\cal E}^{\rm inc}_{\nu}$, and its signs correspond to enhanced ($+$) or weakened ($-$) SPPF after light scattering 
with G-Ps. If both $q_x$ and $x_3$ are large, such scattering is largely suppressed, leaving only a large and flat basin in the upper-right corners of Figs.\,\ref{f5}$a$-\ref{f5}$d$. 
If $q_x$ is very small, the photon and SPP radiative modes dominate, and then, ${\rm Re}\{\alpha^{\rm eff}_{11}(q_x,\omega\vert x_3)\}$ remains negative and becomes independent of $x_3$.
When $q_x$ is intermediate, the SPP evanescent modes start entering in with increasing $\omega$ up to $\Omega_{\rm r}$. In this case, 
the positive-peak strength reduces and the peak coverage is squeezed into a smaller $x_3$ region where the localization of SPPF is still insignificant. 
In addition, the positive peak splits into two islands at $\omega=0.9\,\Omega_{\rm r}$ and eventually switches to a negative peak followed by a negative 
constant at $\omega=\Omega_{\rm r}$. On the other hand, when $q_x$ becomes very large for a strongly-localized SPPF, 
its scattering by G-P becomes negligible except for the region very close to the surface as shown by the sharp negative edges in the lower-right corners of Figs.\,\ref{f5}$a$-\ref{f5}$d$.
\medskip

\begin{figure}
\centering
\includegraphics[width=0.65\textwidth]{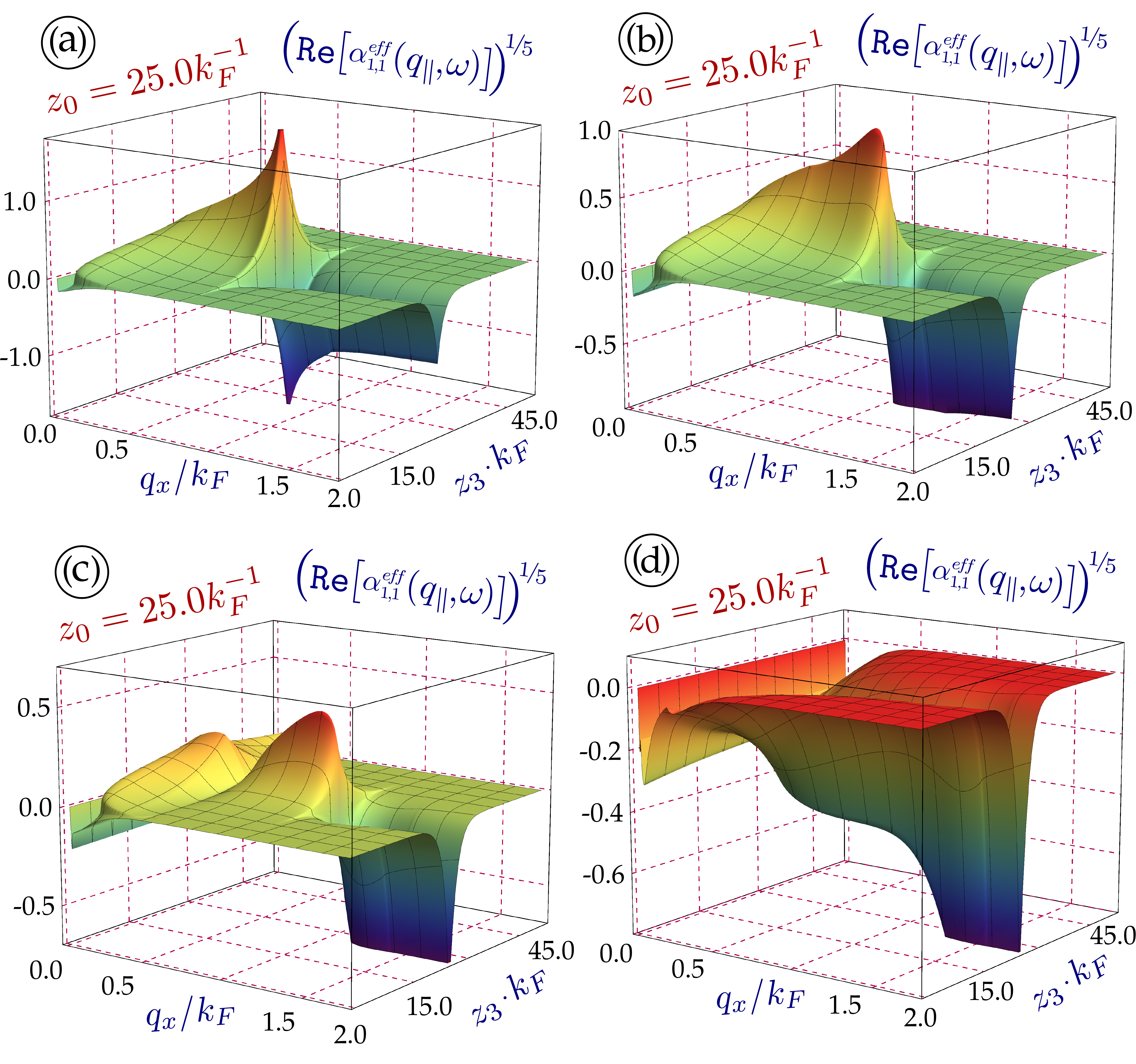}
\caption{3D plots for $[{\rm Re}\{\alpha^{\rm eff}_{11}(q_x,\omega\vert x_3)\}]^{1/5}$ with $\omega/\Omega_r=0.7\,(a),\,0.8\,(b),\,0.9\,(c),\,1.0\,(d)$, where $k_Fz_0 = 25$, $\hbar\Omega_r/E_F=1.78$.}
\label{f6}
\end{figure}

By moving the graphene sheet so it becomes well separated from the conductor surface, as shown in Fig.\,\ref{f6}, we expect the scattering effects from G-Ps on the SPPF to be limited to a narrow area surrounding the graphene sheet 
for large $q_x$ values. Indeed, when $q_x$ is large, we find ${\rm Re}\{\alpha^{\rm eff}_{11}(q_x,\omega\vert x_3)\}=0$ for $x_3$ far away from the graphene sheet at $z_0$ in the upper- and lower-right corners of Figs.\,\ref{f6}$a$-\ref{f6}$d$. For small $q_x$ values, however, the positive peak appears, as in Fig.\,\ref{f5}, and its coverage crawls out along $x_3=z_0$ to a relative large $q_x$ region 
followed by a negative sharp edge, 
although its peak strength decreases with increasing $\omega$ towards $\Omega_r$.
This extended region becomes separated from the positive peak at $\omega=0.9\,\Omega_{\rm r}$ in Fig.\,\ref{f6}$c$ to form an island, 
and both the positive peak and island disappear and are eventually replaced by negative sharp and stepped edges at $\omega=\Omega_{\rm r}$ in Fig.\,\ref{f6}$d$.
\medskip

\noindent
\underline{\em Summary}:\ \ An analogy to Newton's third law in classical mechanics has been demonstrated by the effect of electron back action on the hybridization of radiative and evanescent fields using a retarded interaction, 
which is seen as the hybrid dispersions for both radiative (small $q_x$ range) and evanescent (large $q_x$ range) field modes.
Instead of a reaction force, the back action in this study is an induced optical-polarization field from the Dirac plasmons, 
which resonantly redistributes an incident surface-plasmon-polariton field by scattering.  
The localization characteristics of such a retarded interaction ensures a very high sensitivity to dielectric environments surrounding the graphene, including variations in the conducting substrate, cladding layer,  
electronic properties of embedded graphene, as well as the graphene distance from the conductor surface. 
This provides a unique advantage in wavelength-sensitive optical scrutinizing for chemically-active molecules or proteins bounded with carbon atoms in graphene.  
\medskip

The optical probing tools discussed in this study include either scattering or optical absorption of an incident electromagnetic field. 
For light scattering, we calculate the spatial-temporal dependence of a Fourier transformed scattering matrix, which clearly displays the scattering enhancement, weakening and even suppression 
as functions of both graphene separations ($z_0$) from the conductor surface and wave numbers ($q_x$) of the evanescent surface-plasmon-polariton field at several frequencies close to the localized surface-plasmon resonance.   
This derived scattering matrix lays the foundation for constructing an effective-medium theory commonly employed in finite-difference time-domain methods\,\cite{add17} for solving Maxwell's equations.
For field absorption, the double peaks associated with hybrid surface and graphene plasmon modes at the high-energy side are shown to be dominant for high conductor plasma frequencies. 
However, the round peak at the low-energy side takes the dominant role at low plasma frequencies. Additionally, this round peak shows that 
localized modes can be greatly enhanced when the graphene is moved close to the conductor surface. 
These unique features in resonant absorption enable the selective excitation of radiative polariton modes for their condensation and a threshold-free laser afterwards. 
\medskip

\noindent
\underline{\em Acknowledgments}:\ \ D.H. would like to thank the support from the Air Force Offce of Scientific Research (AFOSR).

\newpage

\begin{references}
\bibitem{add8}H. Haug and S. W. Koch, \textsl{Quantum Theory of the Optical and Electronic Properties of Semiconductors} (Fourth Edition, World Scientific Publishing Co. Pte. Ltd., 2004).

\bibitem{add26}F. Rossi and T. Kuhn, Rev. Mod. Phys. {\bf 74}, 895 (2002).

\bibitem{add9}M. Lindberg and S. W. Koch, Phys. Rev. B {\bf 38}, 3342 (1988).

\bibitem{add29}M. Kira and S. W. Koch, Progress in Quantum Electronics {\bf 30}, 155 (2006).

\bibitem{ref1}D. H. Huang, M. M. Easter, G. Gumbs, A. A. Maradudin, S.-Y. Lin, D. A. Cardimona and X. Zhang, Opt. Expr. {\bf 22}, 27576 (2014).

\bibitem{add12}D. H. Huang, M. M. Easter, G. Gumbs, A. A. Maradudin, S.-Y. Lin, D. A. Cardimona, and X. Zhang, Appl. Phys. Lett. {\bf 104}, 251103 (2014).

\bibitem{add1}K. Novoselov, A. K. Geim, S. Morozov, D. Jiang, M. Katsnelson, I. Grigorieva, S. Dubonos, and A. Firsov, Nature {\bf 438}, 197 (2005).

\bibitem{add2}A. K. Geim and K. S. Novoselov, Nature Materials {\bf 6}, 183 (2007).

\bibitem{add3}Y. Zhang, Y.-W. Tan, H. L. St\"ormer, and P. Kim, Nature {\bf 438}, 201 (2005).

\bibitem{add4}A. K. Geim, Science {\bf 324}, 1530 (2009).

\bibitem{oe1}S. Christopoulos, G. B. H. von H\"ogersthal, A. J. D. Grundy, P. G. Lagoudakis, A.V. Kavokin, J. J. Baumberg, G. Christmann, R. Butt\'e, E. Feltin, J.-F. Carlin, and N. Grandjean, 
Phys. Rev. Lett. {\bf 98}, 126405 (2007).

\bibitem{oe2}S. I. Tsintzos, N. T. Pelekanos, G. Konstantinidis, Z. Hatzopoulos, and P. G. Savvidis, Nat. Lett. {\bf 453}, 372 (2008).

\bibitem{oe3}P. Bhattacharya, B. Xiao, A. Das, S. Bhowmick, and J. Heo, Phys. Rev. Lett. {\bf 110}, 206403 (2013).

\bibitem{oe4}C. Schneider, A. Rahimi-Iman, N. Y. Kim, J. Fischer, I. G. Savenko, M. Amthor, M. Lermer, A. Wolf, L. Worschech, V. D. Kulakovskii, I. A. Shelykh, M. Kamp, S. Reitzenstein, A. Forchel, Y. Yamamoto, 
and S. H\"ofling, Nat. {\bf 497}, 348 (2013).

\bibitem{add28}E. L. Albuquerque and M. G. Cottam, Phys. Rep. {\bf 233}, 67 (1993).

\bibitem{add5}B. Wang, X. Zhang, F. J. Garc\'ia-Vidal, X. Yuan, and J. Teng, Phys. Rev. Lett. {\bf 109}, 073901 (2012).

\bibitem{add6}M. Liu, X. Yin, and X. Zhang, Nano Letters {\bf 12}, 1482 (2012).

\bibitem{add7}M. Liu, X. Yin, E. Ulin-Avila, B. Geng, T. Zentgraf, L. Ju, F. Wang, and X. Zhang, Nature {\bf 474}, 64 (2011).

\bibitem{add10}F. Koppens, T. Mueller, P. Avouris, A. Ferrari, M. Vitiello, and M. Polini, Nature nanotechnology {\bf 9}, 780 (2014).

\bibitem{add11}A. A. Maradudin and D. L. Mills, Phys. Rev. B {\bf 11}, 1392 (1975).

\bibitem{add15}M. G. Cottam and A. A. Maradudin, ``Surface linear response functions'', in \textsl{Surface Excitations}, eds. V. M. Agranovich and R. Loudon (North-Holland, Amsterdam, 1984), pp. 1-194.

\bibitem{add27}J. M. Pitarke, V. M. Silkin, E. V. Chulkov, and P. M. Echenique, Rep. Prog. Phys. {\bf 70}, 1 (2007).

\bibitem{add24}A. V. Zayats, I. I. Smolyaninov, and A. A. Maradudin, Phys. Rep. {\bf 408}, 131 (2005).

\bibitem{add14}A. Iurov, G. Gumbs, D. Huang, and V. Silkin, Phys. Rev. B {\bf 93}, 035404 (2016).

\bibitem{add18}M. S. Tame, K. R. McEnery, S. K. \"Ozdemir, J. Lee, S. A. Maier, and M. S. Kim, Nat. Phys. {\bf 9}, 329 (2013).

\bibitem{add25}F. de Le\'on-P\'erez, G. Brucoli, F. J. García-Vidal, and L. Mart\'in-Moreno, New J. Phys. {\bf 10}, 105017 (2008). 

\bibitem{add31}A. N. Grigorenko, M. Polini, and K. S. Novoselov, Nature Photonics {\bf 6}, 749 (2012).

\bibitem{add32}D. N. Basov, M. M. Fogler, A. Lanzara, F. Wang Y. Zhang, Rev. Mod. Phys. {\bf 86}, 959 (2014).

\bibitem{add19}D. S. Saxon, Phys. Rev. {\bf 100}, 1771 (1955).

\bibitem{add30}F. J. Garc\'ia de Abajo, Rev. Mod. Phys. {\bf 79}, 1267 (2007).

\bibitem{add20}J. van Kranendonk and J. E. Sipe, ``Foundations of the macroscopic electromagnetic theory of dielectric media'', in \textsl{Progress in Optics XV}, ed. E. Wolf (New York: North-Holland, 1977), Chap. 5.

\bibitem{add21}G. D. Mahan and G. Obermair, Phys. Rev. {\bf 183}, 834 (1969).

\bibitem{add22}J. Sipe and J. van Kranendonk, Phys. Rev. A {\bf 9}, 1806 (1974).

\bibitem{add23}W. Lamb, D. M. Wood, N. W. Ashcroft, Phys. Rev. B {\bf 21}, 2248 (1980).

\bibitem{add13}M. Wojcik, M. Hauser, W. Li, S. Moon, and K. Xu, Nature Communications {\bf 6}, 7384 (2015).

\bibitem{ref2}G. Gumbs and D. H. Huang,
\textsl{Properties of Interacting Low-Dimensional Systems} (John Wiley \& Sons, 2011), Chap. 2.

\bibitem{ref3}D. H. Huang, G. Gumbs and O. Roslyak,
Applied Optics {\bf 52}, 755 (2013).

\bibitem{add16}O. Roslyak, G. Gumbs and D. H. Huang, J. Appl. Phys. {\bf 109}, 113721 (2011).

\bibitem{lorentz}D. H. Huang and Y. Zhao, Physical Review A {\bf 51}, 1617 (1995).

\bibitem{add17}A. F. Oskooi, D. Roundy, M. Ibanescu, P. Bermel, J. D. Joannopoulos, S. G. Johnson, Computer Physics Communications {\bf 181}, 687 (2010).
\end{references}
\end{document}